\documentclass[11pt]{iopart}
\usepackage[T1]{fontenc}
\usepackage{iopams} 
\usepackage{graphicx}
\usepackage[colorlinks=true,linkcolor=blue,citecolor=red,urlcolor=magenta]{hyperref}
\usepackage[dvipsnames]{xcolor}
\usepackage{charter} 
 
\begin{document} 

\title[Variational ansatz for $p$-wave fermions]{Variational ansatz for $p$-wave fermions \\ confined in a one-dimensional harmonic trap}
\author{Przemys\l aw Ko\'scik}
\address{Department of Computer Sciences, State Higher Vocational School in
Tarn\'{o}w \\ ul. Mickiewicza 8, PL-33100 Tarn\'{o}w, Poland}
\author{Tomasz Sowi\'nski}
\address{Institute of Physics, Polish Academy of Sciences \\ Aleja Lotnikow 32/46, PL-02668 Warsaw, Poland}

\begin{abstract}
We propose a very accurate and efficient variational scheme for the ground state of the system of $p$-wave attractively interacting fermions confined in a one-dimensional harmonic trap. By the construction, the method takes the non-analytical part of interactions exactly into account and thus it approximates the true ground-state wave function in a whole range of interactions very accurately. Within the method, we determine different properties of the system for a different number of particles and different interactions. In this way, we explore how the system and its features transit from the ideal non-interacting Fermi gas to the system of infinitely strong attractions. Additionally, we demonstrate that the ansatz may also be used on a repulsive branch of interactions where other numerical methods break down. The presented method of including zero-range interactions is very universal and may be easily generalized to other one-dimensional confinements.
\end{abstract}

\section{Introduction}
Appropriate description of strongly correlated quantum many-body systems offering an adequate explanation of their different measurable properties is one of the most challenging tasks for theoretical physics from over 60 years \cite{1971FetterBook}. Besides theoretical reasons, it became fundamentally important recently due to tremendous progress in quantum engineering giving opportunities to coherent control of matter and light on atomic scales where an accurate theoretical description is required. The fundamental obstacle for all the straightforward descriptions of many-body systems originates in the mathematical complexity of a many-body Schr\"odinger equation which, in fact, can be analytically solved only in several specific cases. To the most famous examples belong: the Moshinsky model and its variations \cite{1968MoshinskyAJP,1985BialynickiLetMPhys,2000ZaluskaPRA,2013KoscikFBS,2017KlaimanChemPhys}, the Lieb-Liniger model \cite{1963LiebPR,1963LiebPRb}, or the Calogero-Sutherland model \cite{1971CalogeroJMP,1971SutherlandJMP}. Even in the case of only two interacting particles the list is not significantly extended and contains only few additional solutions: the famous Busch {\it et al.} solution for $s$-wave contact forces \cite{1998BuschFoundPhys}, its variations \cite{2005IdziaszekPRA,2008LiangPhysScripta,2020ChenARX} and generalization to $p$-wave forces \cite{2006SunPRA}, the Gao solutions for $1/r^6$ and $1/r^3$ potentials \cite{1998GaoPRA,1999GaoPRA}, and specific solutions for finite-range interactions modeled by a step function \cite{2018KoscikSciRep,2019KoscikSciRep}.

The situation becomes even more complicated and almost hopeless when inter-particle interactions are modeled by singular functions. Then, many direct numerical attempts are not able to capture subtle features of such interactions and simply break down. One of the simplest, but still realistic models dominated by these kinds of problems is the one-dimensional model of $N$ identical spinless fermions interacting via two-body zero-range forces \cite{2004GirardeauPRA,2004KanjilalPRA,2013DiazNatureComm}. In contrast to bosonic systems, in this case, the Pauli exclusion principle precludes any scattering in the $s$-wave channel and the first non-vanishing contribution to interactions comes from the $p$-wave scattering \cite{1999CheonPRL,2003SenJPhysA}. In one-dimensional case this interaction acts highly counterintuitively: if $\phi_1(x)$ and $\phi_2(x)$ are two different wave functions describing relative motion of two fermions then the matrix element of $p$-wave interaction is proportional to the product of their spatial derivatives at the origin, $\propto\partial_x\phi_1|_{x=0}\,\partial_x\phi_2|_{x=0}$.

Bypassing this difficulty in any numerical treatment is not an easy task if one does not take into account the mentioned singularity of inter-particle interactions exactly. Importantly, special care on this problem needs to be put when the number of particles is not large since then any inaccurate approximation may lead to significant discrepancies in physical predictions. It is known that one of the possible paths to overcome this kind of difficulties is to perform approximate calculations with appropriately tailored trial functions \cite{2015GudymaPRA,2016DecampPRA,2016MatveevaNJP,2016AndersenSciRep,2017PecakPRA,2017LangEPJST,2018RizziPRA}. In this work, based on our previous experience with zero-range forces in the bosonic case \cite{2018KoscikEPL}, we propose a very accurate and very efficient way to find approximate ground-state wave function of a few $p$-wave interacting fermions. To show that the method proposed is indeed profitable, we focus on the generic problem of $N$ identical fermions attractively interacting via $p$-wave forces confined in a one-dimensional harmonic trap. Preliminarily, the model has been studied already, mostly in the limit of infinite attractions \cite{2006MinguzziPRA,2014ZhangPRA,2016CuiPRA,2016YangPRAc}. We show that within the framework of our approach we can determine with high accuracy not only the ground-state energy and single-particle properties of the system (determined previously for other scenarios \cite{2005BenderPRL,2007HaoPRA,2016HuPRA}) but also we can go much further and study the inter-particle correlations in position and momentum domain. As an example, we display two-particle probability densities highlighting intriguing correlations, especially in the momentum domain. Importantly, the variational method proposed opens a route to capture different properties of the system for intermediate interactions and therefore to observe their evolution when interactions are tuned along a whole range. Additionally, we show that the method can be also utilized to study properties of the system with repulsive interactions which are challenging for other computational techniques.

\section{The system}
In our work we consider the system of $N$ identical spinless fermions of mass $m$ confined in a one-dimensional harmonic trap of frequency $\omega$ and interacting with zero-range interactions. For convenience, in the following considerations we will express all quantities in units of the harmonic oscillator, {\it i.e.}, all energies, lengths, and momenta will be expressed in units of $\hbar\omega$, $\sqrt{\hbar/m\omega}$, and $\sqrt{\hbar m \omega}$, respectively. As already mentioned in the introduction, the $s$-wave scattering between particles is not present and the first non-vanishing contribution to interactions comes from the zero-range $p$-wave interactions. In a one-dimensional geometry, the interaction potential can be expressed formally as \cite{1986SebaRMP,2003SenJPhysA,2004GirardeauPRA,2019SowinskiRPP}
\begin{equation} \label{pwavepot}
V_{p}(x)=-\frac{g_F}{2}\overleftarrow{\frac{\partial}{\partial x}}\delta(x)\overrightarrow{\frac{\partial}{\partial x}},
\end{equation}
where $x$ is a relative distance between particles and $g_F$ is the effective $p$-wave interaction strength. Interpretation of directed derivatives  in (\ref{pwavepot}) (indicated by arrows) is operational, {\it i.e.}, if $\phi_1(x)$ and $\phi_2(x)$ represent wave functions of quantum states of a relative motion of two fermions then the matrix element of the interaction between these two states is calculated as
\begin{equation}
\langle\phi_1|V_p|\phi_2\rangle = -\frac{g_F}{2} \left.\frac{\mathrm{d} \phi_1^*(x)}{\mathrm{d} x}\right|_{x=0} \left.\frac{\mathrm{d} \phi_2(x)}{\mathrm{d} x}\right|_{x=0}.  
\end{equation}
Taking this into account, the many-body Hamiltonian of the system studied has a form
\begin{equation} \label{Hamiltonian}
{\cal H}=\sum_{i=1}^N\left[-\frac{1}{2}\frac{\partial^2}{\partial x_{i}^2}+\frac{1}{2} x_{i}^2+\sum_{j>i}^N V_{p}(x_{i}-x_{j})\right].
\end{equation}
It turns out that in the one-dimensional scenario the fermionic Hamiltonian (\ref{Hamiltonian}) is exactly equivalent to the problem described by the non-interacting Hamiltonian
\begin{equation}
{\cal H}=\sum_{i=1}^N\left[-\frac{1}{2}\frac{\partial^2}{\partial x_{i}^2}+\frac{1}{2} x_{i}^2\right],
\end{equation}
provided that the many-body eigenstate wave function $\psi(x_1,\ldots,x_N)$ is antisymmetric under an exchange of any two positions and it additionally supports the contact condition for each pair of fermions of the form \cite{2004KanjilalPRA,2004GirardeauOpticsCom,2007HaoPRA}
\begin{equation}\label{condition}
\psi(x_i-x_j=0^+)=-g_F{\partial \over \partial x_{ij}}\psi(x_i=x_j\pm 0),
\end{equation}
where $x_{ij}=x_i-x_j$.
This equivalence simply means that a whole effect of $p$-wave interactions between particles is encoded directly in the condition (\ref{condition}). Our aim is to find a convenient approximate form for the ground-state wave function fulfilling this condition {\it exactly}.

\section{Variational approach} \label{SectionVariational}
Before we present our construction of the trial function for the $N$-particle system, let us note that the problem studied has known analytical solutions in the case of $N=2$ fermions \cite{2006SunPRA}. The eigenenergies and corresponding eigenvectors are found analogously as in the celebrated Busch {\it et. al} problem of two $s$-wave interacting bosons \cite{1998BuschFoundPhys,2006PatilEJP,2009WeiIJMPB,2010SowinskiPRA}. It turns out that for $p$-wave fermions the two-particle ground-state of the Hamiltonian (\ref{Hamiltonian}) is expressed in terms of the confluent hyperbolic function $\textbf{U}\left(a,b,z\right)$ as
\begin{equation} \label{Function2part}
\psi(x_{1},x_{2})=(x_1-x_2)\,\mathrm{e}^{-(x_1^2+x_2^2)/2}\,
 \textbf{U}\left(\frac{3-2\epsilon}{4};\frac{3}{2};\frac{({x_{1}}-{x_{2}})^2}{2}
 \right),
\end{equation}
where the parameter $\epsilon$ is determined by the contact condition (\ref{condition}). Namely, for given interaction strength $g_F$ the wave function (\ref{Function2part}) is the two-particle ground state of the Hamiltonian (\ref{Hamiltonian}) provided that parameter $\epsilon$ is the smallest solutions of the following transcendental equation:
\begin{equation} \label{BuschRelation}
\sqrt{2}\,\Gamma\!\left(\frac{1}{4}-\frac{{\epsilon}}{2}\right)=2 g_F\,\Gamma\!\left(\frac{3}{4}-\frac{{\epsilon}}{2}\right).
\end{equation}
Then, the two-particle ground-state energy is equal to $E=\epsilon + 1/2$.

The exact solution of the problem with more than two interacting fermions is known only in the two extreme limits, {\it i.e.}, the non-interacting case ($g_F=0$) and infinitely strong attractions ($g_F=-\infty$). It should be noted however, that for any number of particles $N$ and for any attractive interaction strength $g_F$ there exists a direct and rigorous mapping of the $p$-wave fermions ground-state wave function $\psi(x_1,\ldots,x_N)$ to the ground-state wave function of $s$-wave repulsively interacting bosons $\Phi(x_1,\ldots,x_N)$ described by the many-body Hamiltonian of the form \cite{2004GirardeauPRA,2010MuthPRA,2016YangPRA}: 
\begin{equation} \label{HamiltonianBosons}
{\cal H}=\sum_{i=1}^N\left[-\frac{1}{2}\frac{\partial^2}{\partial x_{i}^2}+\frac{1}{2} x_{i}^2+g_B\sum_{j>i}^N \delta(x_{i}-x_{j})\right],
\end{equation}
where $g_B$ is the effective $s$-wave interaction strength between bosons. Let us formulate the mapping more precisely. If $\Phi(x_1,\ldots,x_N)$ is the ground-state wave function of the Hamiltonian (\ref{HamiltonianBosons}), then the wave function defined as 
\begin{equation} \label{BFmapping}
\psi(x_1,\ldots,x_N) =\prod_{i<j}\mathrm{sgn}(x_{i}-x_{j})\Phi(x_1,\ldots,x_N),
\end{equation}
is the exact ground-state wave function of the Hamiltonian (\ref{Hamiltonian}) describing $p$-wave fermions (having the same ground-state energy) provided that the corresponding interaction strengths $g_B$ and $g_F$ fulfill the condition $g_B \cdot g_F = -2$. This mapping between $p$-wave fermions and $s$-wave bosons can be viewed as a specific and wide generalization of the famous Bose-Fermi mapping between hard-core bosons and non-interacting fermions formulated in 1960 by Girardeau \cite{1960GirardeauJMP}. Consequently, from the principal point of view, the question of finding the ground state of $p$-wave fermions addressed here is rigorously equivalent to another problem of interacting bosons. However, the problem of efficient construction of the wave function remains unsolved since the exact form of bosonic wave function is not known. Therefore, is it still interesting to find accurate and efficient method of constructing many-body wave function for interacting $p$-wave fermions.

One of natural methods of finding the many-body ground-state wave function is to propose a reasonable family of variational trial functions appropriately tailored to capture the most important features of the system. Having in hand the exact two-body solution (\ref{Function2part}) which is valid for any interaction strength, one can introduce the Jastrow-like trial function for the problem of $N$ particles \cite{1955Jastrow}. In this approximation it is assumed that the most prominent part of inter-particle correlations are captured in the two-body sector described by the two-body solution. Therefore, the trial wave function has a form:
\begin{equation}\label{Jastrow}
\Upsilon_\alpha(x_1,\ldots,x_N) = \mathrm{e}^{-\frac{1}{2}\sum_{i=1}^{N} x_{i}^{2}}\prod_{i<j} \varphi_\alpha(x_{i}-x_{j})
\end{equation}
with a rescaled two-body solution
\begin{equation}
\varphi_\alpha(x) =x\,
 \textbf{U}\left(\frac{3-2\epsilon_{\alpha}}{4};\frac{3}{2};\frac{\alpha^2 x^2}{2}
 \right).
\end{equation}
In this approach, $\alpha$ is the variational parameter of the family and $\epsilon_{\alpha}$ are determined by solving the equation
\begin{equation} \label{BuschRelation1}
\sqrt{2}\,\Gamma\!\left(1/4-{\epsilon_{\alpha}}/{2}\right)=2 g_F\alpha\,\Gamma\!\left(3/4-{\epsilon_{\alpha}}/{2}\right).
\end{equation}
 In principle, by minimizing the energy functional
\begin{equation}
E[\Upsilon_\alpha]= \frac{\langle\Upsilon_\alpha|{\cal H}|\Upsilon_\alpha\rangle}{\langle\Upsilon_\alpha|\Upsilon_\alpha\rangle}
\end{equation}
one can obtain approximation to the ground-state of interacting system of $N$ fermions and corresponding energy. However, in practice, this approach is very demanding due to a quite large computational complexity. This complexity originates in a tangled definition of the rescaled two-body solution $\psi_\alpha(x_1,x_2)$ through the hypergeometric function determined by $\epsilon_\alpha$ being a solution of transcendental equation (\ref{BuschRelation1}). Consequently, the method cannot be easily used for large number of particles. A very similar problem was revealed recently in the case of interacting bosons \cite{2018KoscikEPL}.

To overcome the difficulty described above, instead of $\Upsilon_\alpha(x_1,\ldots,x_N)$, we propose to use different variational ansatz which turns out to be reasonably accurate and numerically very efficient. This approach is still based on the assumption that the dominant part of correlations has two-body origins and can be written as
\begin{equation}\label{jastrow}
\psi_\alpha(x_1,\ldots,x_N)=\mathrm{e}^{-\frac{1}{2}\sum_{i=1}^{N} {x_{i}^{2}}}\prod_{i<j} \phi_{\alpha}(x_{i}-x_{j}).
\end{equation}
However, the correlated pair function $\phi_{\alpha}(x)$ is significantly simplified to the form
\begin{equation}\label{fun}
\phi_{\alpha} (x)=\mathrm{sgn}(x)\left(1-\frac{\mathrm{e}^{-\alpha |x|}}{1-\alpha g_F}\right),
\end{equation}
where $\alpha$ plays a role of the variational parameter. Importantly, the two-body trial function $\phi_\alpha(x)$ has appropriate properties in the vicinity of $x=0$:
\numparts
\begin{eqnarray}
\phi_\alpha(0^{+})&=-\phi_\alpha(0^{-}), \\
\left.\frac{\partial}{\partial{x}}\phi_\alpha(x)\right|_{x\rightarrow 0^{+}}&=\left.\frac{\partial}{\partial{x}}\phi_\alpha(x)\right|_{x\rightarrow 0^{-}}.
\end{eqnarray}
\endnumparts
Thus, the whole variational wave function (\ref{jastrow}) fulfills automatically the discontinuity condition (\ref{condition}) and therefore it includes exactly all effects of inter-particle interactions. Moreover, the ansatz reproduces rigorously the system's ground state in the two extreme limits for any number of particles $N$. First, in the limit of non-interacting fermions ($g_F\rightarrow 0$) the parameter $\alpha\rightarrow 0$ and as a result $\phi_{\alpha}(x)\rightarrow \alpha\,\mathrm{sgn}(x) |x|=\alpha x$. Second, in the fermionic Tonks-Girardeau limit ($g_F\rightarrow -\infty$), one finds $\phi_{\alpha} (x)\rightarrow \mathrm{sgn}(x)$.

At this point we want to stress that the variational ansatz for $p$-wave fermions presented here, due to the mapping (\ref{BFmapping}), is a direct and rigorous consequence of the ansatz provided previously for $s$-wave bosons \cite{2018KoscikEPL}. However, since the mapping is highly non-trivial, it does not mean that all the properties of the fermionic system can be easily deduced from the pure bosonic many-body wave function. We aim to show that the non-obvious form of the Jastrow ansatz (\ref{Jastrow}), which disregards an existence of the exact solution of corresponding two-body problem, is appropriate to explore different and highly non-trivial properties of $p$-wave fermions being elusive for other computational techniques.

\section{The results}
\begin{figure}
\centering
\includegraphics[width=0.7\linewidth]{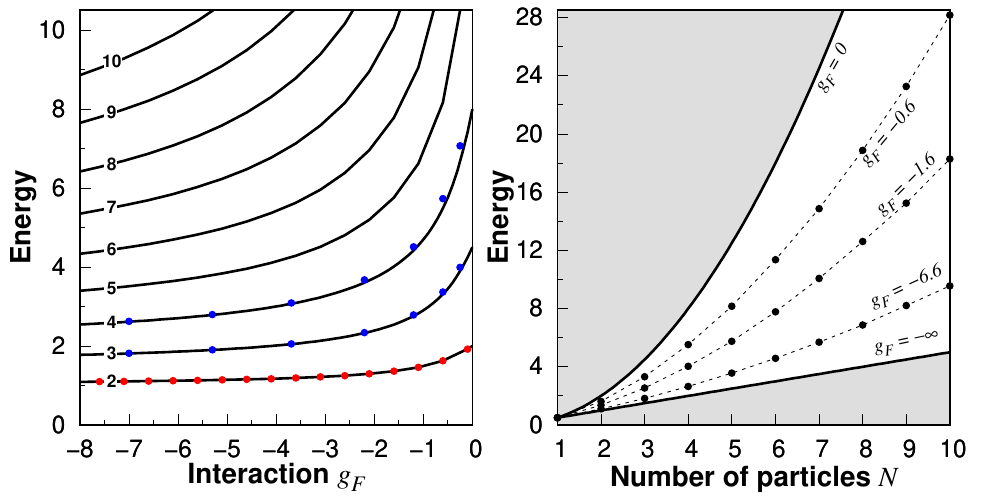}
\caption{Ground-state energy of interacting $p$-wave fermions obtained from the variational approach. {\bf (left panel)} Energy as a function of inter-particle interactions $g_F$ for given number of particles. For comparison, with red dots we mark the exact ground-state energies for $N=2$ calculated analytically as presented in \cite{2006SunPRA}. Blue dots denote corresponding ground-state energies obtained numerically via exact diagonalization. {\bf (right panel)} Energy as a function of the number of fermions $N$ for fixed interactions $g_F$. Note that in the limiting cases $g_F=0$ and $g_F\rightarrow-\infty$ energies are given by $E_0(N)=N^2/2$ and $E_{-\infty}(N)=N/2$, respectively. \label{Fig1}}
\end{figure}

\subsection{Ground-state energy}

The variational trial wave function (\ref{jastrow}) is a very efficient tool for determining different properties of many interacting $p$-wave fermions for intermediate interaction strengths. First, it allows us to find quickly the ground-state energy for a given number of particles $N$. As shown in Fig.~\ref{Fig1}, ground-state energies obtained in this way nicely interpolate between two extreme points -- the non-interacting energy $E_{0}(N)=N^2/2$ and the ground-state energy in the limit of infinite attractions $E_{-\infty}(N)=N/2$. Moreover, the energies almost ideally match the exact values for $N=2$ obtained from relation (\ref{BuschRelation}) (red dots) and agree also with ground-state energies obtained with the optimized exact diagonalization method \cite{2018KoscikPLA,2020KoscikFewBody} of the many-body Hamiltonian for $N=3$ and $N=4$ (blue dots). Interestingly, it should be noted that for small attractions and larger number of particles energies obtained via exact diagonalization are slightly higher than that obtained with the ansatz. This tiny discrepancy is caused by high inaccuracy of diagonalization method in the limit of vanishing $p$-wave interactions (which corresponds to strong $s$-wave bosonic repulsions). The variational ansatz works considerably better in this range of interactions. To show that the method can be successfully used for a quite large number of particles, in the right panel of Fig.~\ref{Fig1} we plot the ground-state energy as a function of the number of particles for some particular values of interactions.

\subsection{One-body properties}

\begin{figure}
\centering
\includegraphics[width=0.7\linewidth]{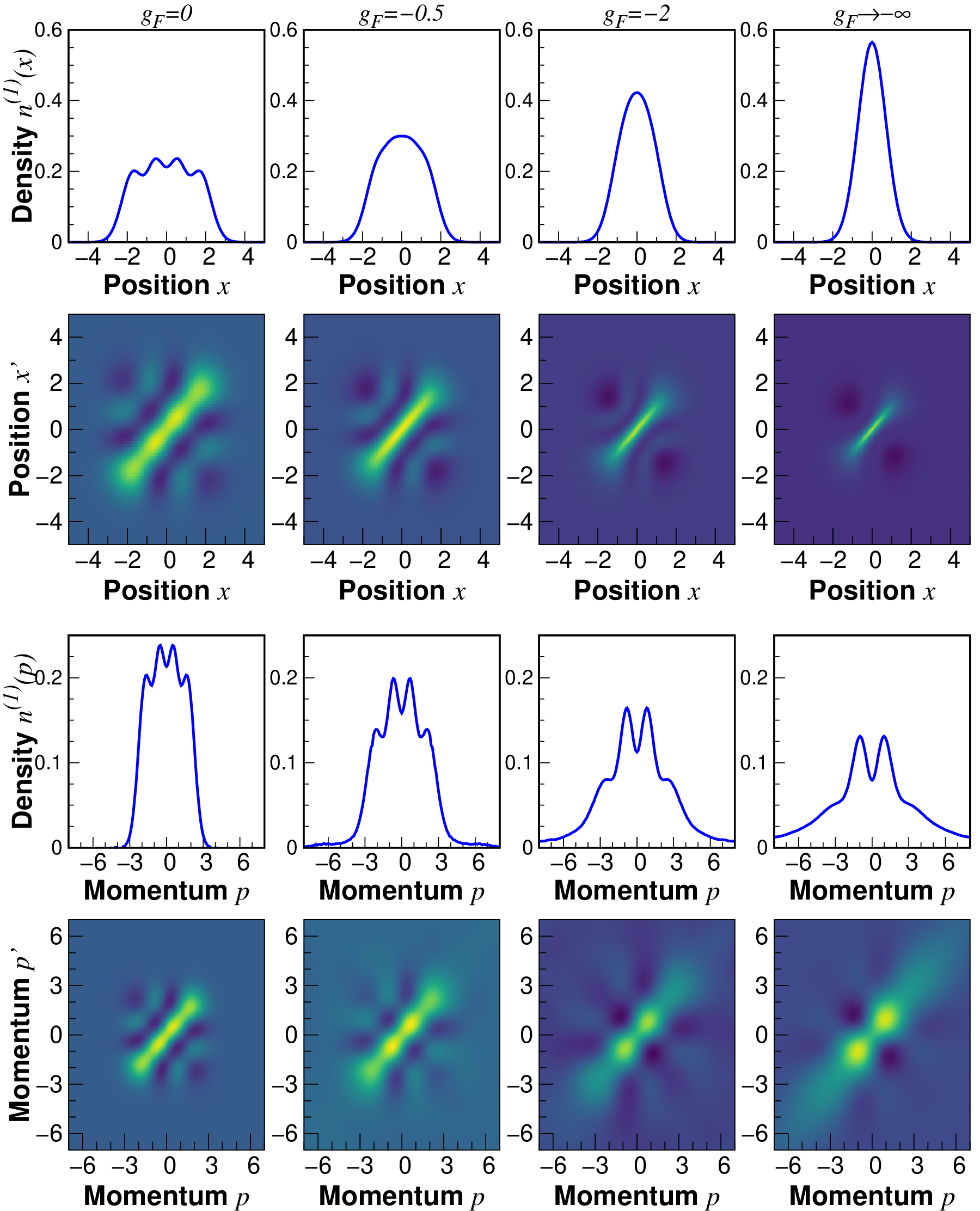}
\caption{Single-particle properties of the system of $N=4$ fermions and different interaction strengths. First two rows correspond to the single-particle density profile $n^{(1)}(x)$ and the single-particle reduced density matrix $\rho^{(1)}(x,x^\prime)$ in the position representation, while next two rows to the corresponding quantities $\tilde{n}^{(1)}(p)$ and $\tilde\rho^{(1)}(p,p^\prime)$ in the momentum representation (the momentum distribution and the single-particle reduced density matrix in momentum domain, respectively). Corresponding results for $N=3$ and $N=5$ are displayed in Fig.~\ref{FigA1} and Fig.~\ref{FigA2}, respectively. \label{Fig2}}
\end{figure}

When the variational ground-state energy is found one automatically has an approximate representation of the many-body ground state for interacting $p$-wave fermions. This enables one to study different properties of the system. All single-particle ones are encoded in the single-particle reduced density matrix. In the position representation it can be calculated straightforwardly from the ground-state wave function as:
\begin{equation} \label{RDM1}
\rho^{(1)}(x,x^{\prime}) = \int\!\mathrm{d}x_{2}\ldots\mathrm{d}x_{N}\,\psi^*(x,x_{2},\ldots,x_{N})\psi(x^{\prime},x_{2},\ldots,x_{N}).
\end{equation}
Its diagonal part $n^{(1)}(x)=\rho^{(1)}(x,x)$ encodes the single-particle density profile. Taking the case of $N=4$ particles as an instructive example, in the first two rows in Fig.~\ref{Fig2} we plot these quantities for different interactions. It is clearly seen that along with increasing $p$-wave attractions, the density profile and single-particle density matrix change their shapes. In the limit of infinite attractions ($g_F=-\infty$) both of them resemble corresponding features of non-interacting bosons which is in a full accordance with the mapping mentioned above.

The situation changes significantly when, instead of spatial, momentum properties of the system are discussed, since momentum distributions cannot be easily determined by simple mapping from the associated $s$-wave bosonic system. This unfeasibility originates in a simple fact that the mapping between attractive fermions and repulsive bosons is a knotty transformation in the position domain.

The simplest properties of the system in the momentum domain are encoded in the single-particle density matrix defined as
\begin{equation} \label{RDM1momentum}
\tilde{\rho}^{(1)}(p,p')= {1\over 2\pi}\int \rho^{(1)}(x,x^{\prime})\mathrm{e}^{-ipx}\mathrm{e}^{ipx'}\mathrm{d}x\mathrm{d}x^{\prime}
\end{equation}
and its diagonal part (the single-particle momentum distribution) of the form
\begin{equation}
\tilde{n}^{(1)}(p)= {1\over 2\pi}\int \rho^{(1)}(x,x^{\prime})e^{-ip (x-x^{\prime})}\mathrm{d}x\mathrm{d}x^{\prime}.
\end{equation}
In two bottom rows in Fig.~\ref{Fig2}, we display these distributions for $N=4$ and corresponding interactions. It turns out that along with increasing attractions in the system higher momenta of single fermions are accessible. Thank to the method used, now one has an access to the ground-state distributions also for intermediate interactions. In this way, one can easily observe how the two exterior peaks present in the distribution of non-interacting system are smeared along with increasing attractions while remaining two are enhanced. From this point of view, it is also very instructive to compare this behavior for different number of particles. Therefore, in the Appendix we present them for another cases with $N=3$ and $N=5$ particles (see Fig.~\ref{FigA1} and Fig.~\ref{FigA2}, respectively). As it is seen, in these cases (note odd number of particles) the central peak is enhanced while only two the most external are smeared.

\begin{figure}
\centering
\includegraphics[width=0.7\linewidth]{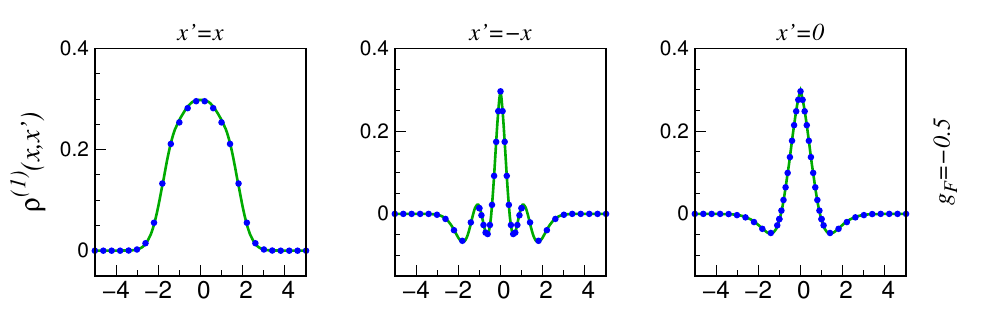}
\includegraphics[width=0.7\linewidth]{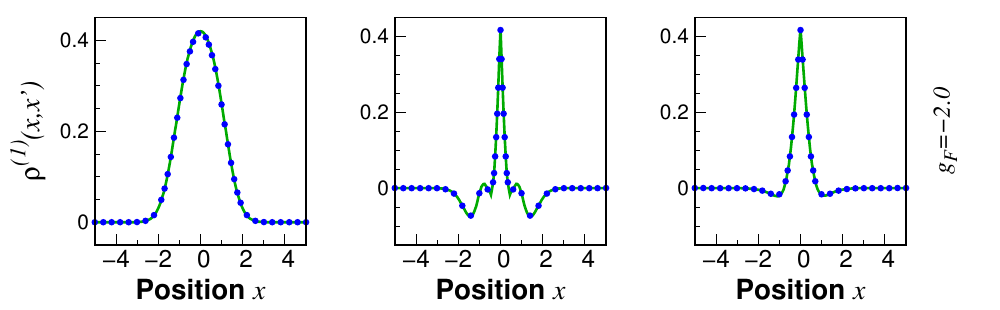}
\caption{Comparison of the single-particle reduced density matrices $\rho^{(1)}(x,x')$ obtained for $N=4$ fermions from the variational ansatz approach (solid green lines) and obtained numerically via exact diagonalization (blue dots). Successive columns correspond to cuts of the density matrix along three different directions: diagonal ($x'=x$), off-diagonal ($x'=-x$), and horizontal ($x'=0$). Top and bottom  row correspond to two different interaction strengths $g_F=-0.5$ and $g_F=-2.0$, respectively. Corresponding results for $N=3$ are displayed in Fig.~\ref{FigA3}.\label{Fig3}}
\end{figure} 

To make sure that single-particle properties predicted by the variational ansatz are credible, we perform additional cross-check with independent method of a direct numerical diagonalization of the many-body Hamiltonian. In Fig.~\ref{Fig3} we compare values of the reduced density matrix $\rho^{(1)}(x,x')$ obtained with the variational ansatz (green solid lines) and the exact diagonalization (blue dots) along three different cuts: the diagonal line $x=x'$ (corresponding to the density profile $n^{(1)}(x)$), the off-diagonal line $x'=-x$, and the horizontal line $x'=0$. The results are displayed for $N=4$ and two different interaction strengths $g_F=-0.5$ and $g_F=-2.0$. Similar comparison for $N=3$ fermions is presented in Fig.~\ref{FigA3} in the Appendix. It is clear that the results obtained with these two complementary methods are almost ideally compatible in all the cases showed and therefore credibility of the ansatz proposed is significantly amplified.  

It is worth to note that obtaining the position or momentum dependent quantities from the variational ansatz approach are numerically much less demanding than from the exact diagonalization framework. Typically, the diagonalization is done in the Fock basis build from single-particle orbitals of the non-interacting system and therefore the many-body ground state is given as specific decomposition coefficients in this basis. To obtain position or momentum depended quantities, like reduced density matrices (\ref{RDM1}) and (\ref{RDM1momentum}), one needs to perform appropriate and numerically time-consuming summation over a whole Fock basis for each specific grid point. This procedure is exceptionally unproductive when one analyzes rapidly-changing quantities requiring a very dense grid (like density matrices in the momentum domain) or quantities having many dimensions (like higher order correlations). From this point of view the variational ansatz proposed is significantly less demanding and straightforward. 

\begin{figure}
\centering 
\includegraphics[width=0.7\linewidth]{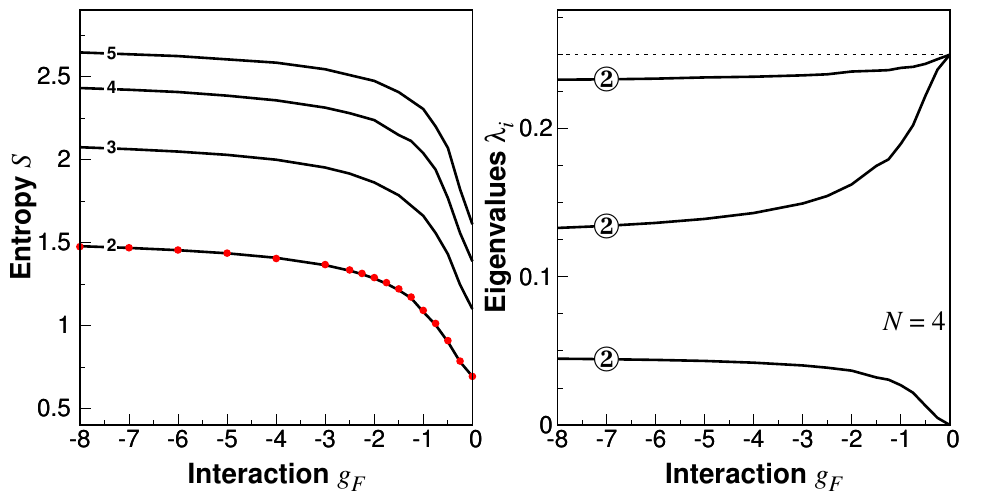}
\includegraphics[width=0.7\linewidth]{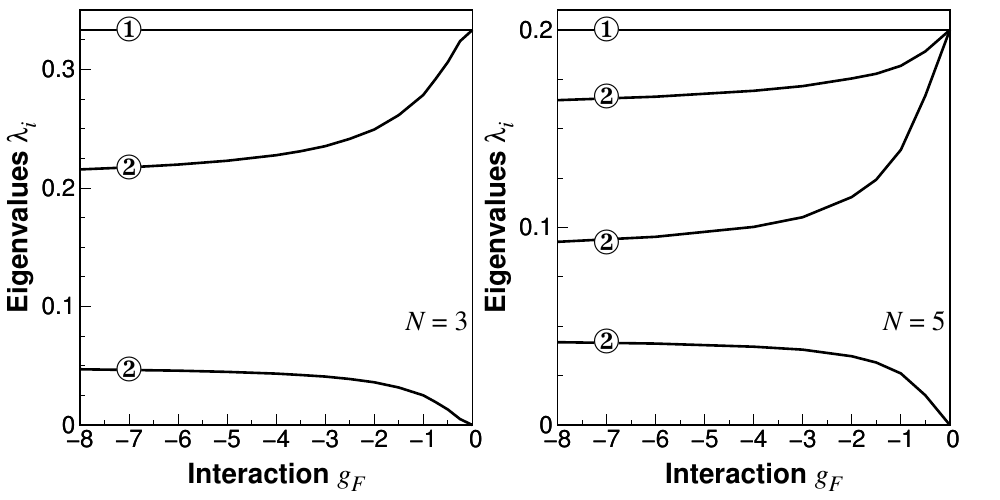}
\caption{Properties of the single-particle reduced density matrix (\ref{Decomp1RDM}) as functions of the interaction strength. {\bf (top left)} The entanglement entropy $S$ as a function of interactions for different number of particles. For comparison, we mark the exact results obtained for $N=2$ by red dots. {\bf (top right and bottom)} Spectrum ($N+2$ the largest eigenvalues) of the single-particle reduced density matrix as a function of interactions for different number of particles $N$. Numbers in circles indicate degeneracy. Dotted line at $\lambda=1/N$ marks the value of $N$-fold degenerated eigenvalues in the non-interacting case.\label{Fig4}}
\end{figure}

At this point let us also mention that a relatively easy access to the full single-particle density matrix of the system provided by the proposed variational scheme gives us also a direct way to quantify non-classical inter-particle correlations in the system. Most simply, this can be done by performing spectral decomposition of the single-particle density matrix
\begin{equation} \label{Decomp1RDM}
\rho^{(1)}(x,x')=\sum \lambda_i \eta^*_i(x)\eta_i(x'),
\end{equation}
  where $\lambda_i$ and functions $\eta_i(x)$ are eigenvalues and corresponding natural orbitals of the single-particle density matrix. Since the system contains $N$ indistinguishable fermions, even in the non-interacting case ($g_F=0$) the decomposition is not trivial and has exactly $N$ non-zero eigenvalues $\lambda_1=\lambda_2=\ldots=\lambda_N$. They correspond to $N$ single-particle orbitals forming the Slater determinant describing the state of the system. For non-vanishing interactions more than $N$ orbitals contribute to the density matrix and the situation becomes more complicated. Nonetheless, still a general structure of the single-particle density matrix is rigorously known and follows directly from the form of the many-body wave function being a product of the symmetric bosonic part and $N-1$ sign-functions of relative positions of particles \cite{1963AndoRevModPhys,1963ColemanRevModPhys}. For the even number of particles $N$ all the eigenvalues \{$\lambda_i\}$ are evenly degenerated. In contrast, for odd $N$, exactly one of the eigenvalues is always equal to $1/N$ and remaining ones are evenly degenerated. As an example, in the top right panel in Fig.~\ref{Fig4} we display the dependence of a few the largest eigenvalues $\lambda$ as functions of interactions for the system of $N=4$ fermions. Corresponding plots for $N=3$ and $N=5$ particles are supplemented in the bottom of the same figure. It is clear that our variational approach appropriately reproduces the structure of the reduced density matrix. Interestingly, close to the non-interacting limit, all the eigenvalues rapidly change their values and for intermediate interaction strengths ($g_F\approx -4$ for $N=4$) saturate at values being very close to their values for infinite attractions. It may suggest that many important one-body features of the system achieved in the limit $g_F\rightarrow-\infty$ are exhibited by the system already for intermediate interactions. Note that even in the limiting case of infinite attractions (corresponding to non-interacting $s$-wave bosons) the spectral structure of the single-particle reduced density matrix is not trivial and substantially different from the corresponding bosonic system. This dissimilarity is a direct manifestation of non-unitarity of the Bose-Fermi mapping procedure (\ref{BFmapping}) which is performed always in the position representation.

Obviously, the exact number of contributing orbitals and corresponding eigenvalues is not determined by general theorems and they depend on the amount of non-trivial correlations present in the system. They can be quantified by the von Neumann entanglement entropy defined as
\begin{equation}
S = -\sum_i \lambda_i \mathrm{ln}\lambda_i.
\end{equation}
The entropy is bounded from below by its value in the non-interacting limit, $S_0=\ln N$. In the right panel in Fig.~\ref{Fig4} we present the dependence of the entanglement entropy $S$ on interactions for a different number of particles. It is clear, that entropy monotonically increases with interaction strength which signals a monotonic increase of correlations in the system. However, exactly as anticipated by the behavior of eigenvalues, the entropy quickly saturates on its value reached in the limit of infinite attractions. Additionally, in the case of $N=2$ fermions, we mark the exact values of the entropy provided by the exact solution (\ref{Function2part}). With this comparison, it is clear that the variational approach proposed appropriately captures also quantitative predictions for non-trivial one-body coherence.

\subsection{Two-body correlations}
\begin{figure}[t]
\centering
\includegraphics[width=0.7\linewidth]{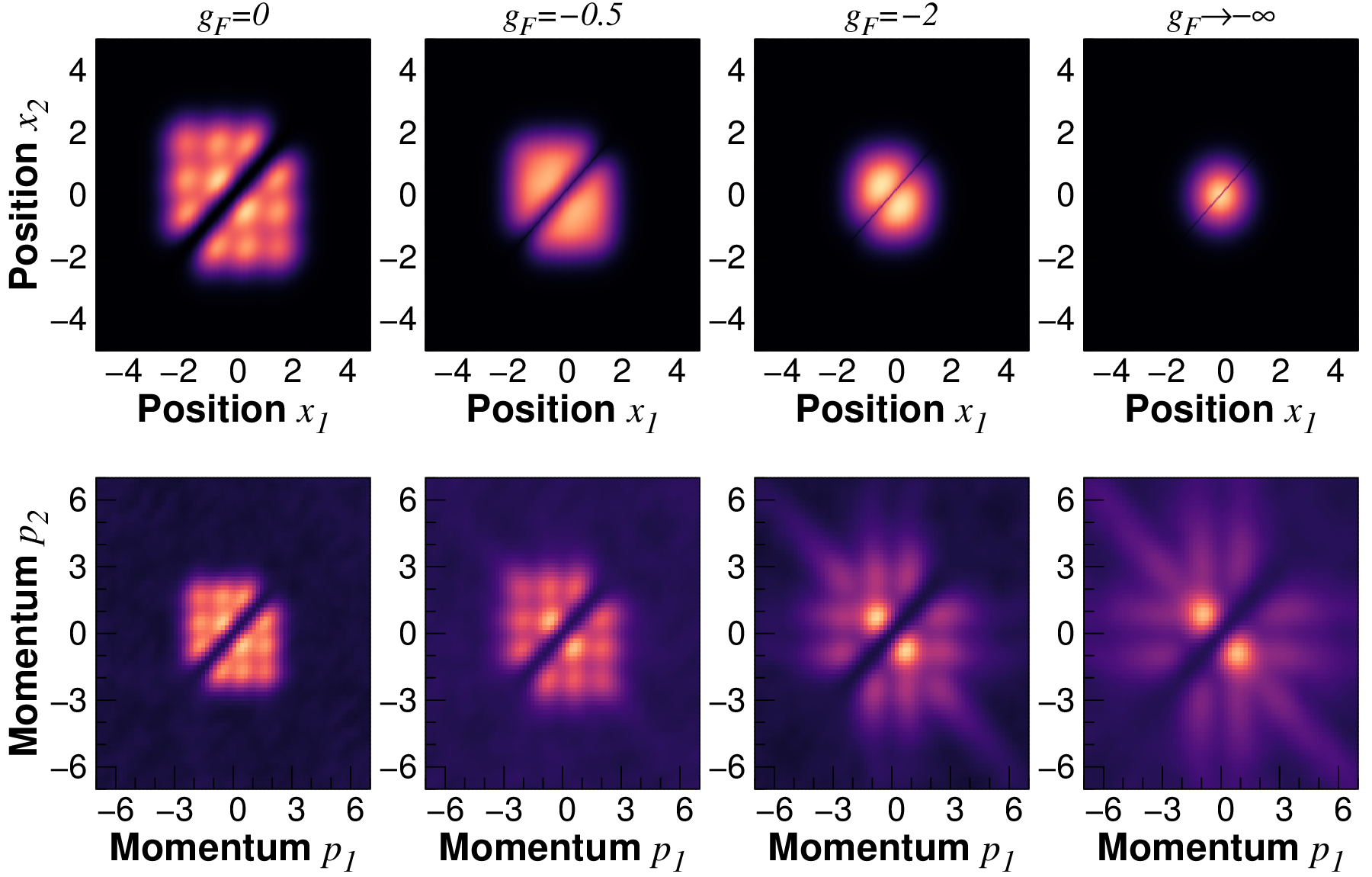}
\caption{Two-particle density profile in position {\bf (top panel)} and momentum representation {\bf (bottom panel)} for $N=4$ fermions and the same interactions as in Fig.~\ref{Fig2}. Corresponding results for $N=3$ and $N=5$ are displayed in Fig.~\ref{FigA1} and Fig.~\ref{FigA2}, respectively.\label{Fig5}}
\end{figure}

The most important advantage of our variational method is its ability to predict higher-order correlations between interacting fermions. Since the full many-body wave function is appropriately represented in a whole range of interactions, it gives a direct view on changes of different non-trivial correlations under tuning interaction strength. For example, one can easily consider two-particle density profiles in the position and momentum domains. They are defined as
\numparts
\begin{eqnarray} 
n^{(2)}(x_{1},x_{2}) &= \rho^{(2)}(x_{1},x_{2},x_{1},x_{2}), \label{2partprofilesa}\\
\tilde{n}^{(2)}(p_1,p_2) &= {1\over 4\pi^2}\int\!\mathrm{d}x\mathrm{d}x^{\prime}\mathrm{d}y\mathrm{d}y^{\prime}\,\rho^{(2)}(x,y,x^{\prime},y^{\prime})\mathrm{e}^{-ip_1 (x-x^{\prime})}\mathrm{e}^{-i p_2 (y-y^{\prime})},\label{2partprofilesb}
\end{eqnarray}
\endnumparts
where we introduced the two-particle reduced density matrix of the form
\begin{equation}
\rho^{(2)}(x,y,x^{\prime},y^{\prime})= \int \!\mathrm{d}x_{3}\ldots \mathrm{d}x_{N}\,\psi^*(x,y,x_{3},\ldots x_{N})\psi(x^{\prime},y^{\prime},x_{3},\ldots,x_{N}).
\end{equation}
Physically, the profiles (\ref{2partprofilesa}) and (\ref{2partprofilesb}) can be interpreted directly as probability densities of finding two fermions with positions ($x_1,x_2$) or momenta ($p_1,p_2$) in a simultaneous measurement of two particles. Therefore, they are the simplest quantities capturing geometrical (in positions or in momenta) arrangement of particles in the many-body ground state. It turns out that properties of these two distributions crucially depend on interactions. To visualize this, in Fig.~\ref{Fig5} we plot them for the system of $N=4$ particles exactly for the same interaction strengths as in Fig.~\ref{Fig2}. Interestingly, the two-particle distribution in the position domain undergoes a smooth transition from the square-like to the circle-like shape when attractive forces are enhanced. This transition is assisted by a significant reduction of the forbidden region along the diagonal $x_1=x_2$. This behavior does not qualitatively depend on the number of particles (see Fig.~\ref{FigA1} and Fig.~\ref{FigA2} in Appendix for $N=3$ and $N=5$, respectively). Contrary, in the momentum space the situation is completely different and it strongly depends on the parity of the number of fermions. While in the non-interacting case, due to the symmetry of canonical variables $x\leftrightarrow p$, the correlation function has also a square-like shape, along with increasing attractions some enhancements of two-body correlations in particular directions appear. Although in the case of $N=4$ particles almost only the ordinary pairing of opposite momenta is supported (clear enhancement along the line $p_1+p_2=0$), for odd number of particles (see Fig.~\ref{FigA1} and Fig.~\ref{FigA2}) we notice additional strong enhancement of pairs in which only one of particles carries all momentum (enhancement along lines $p_1=0$ and $p_2=0$, respectively).

\begin{figure}[t]
\centering
\includegraphics[width=0.7\linewidth]{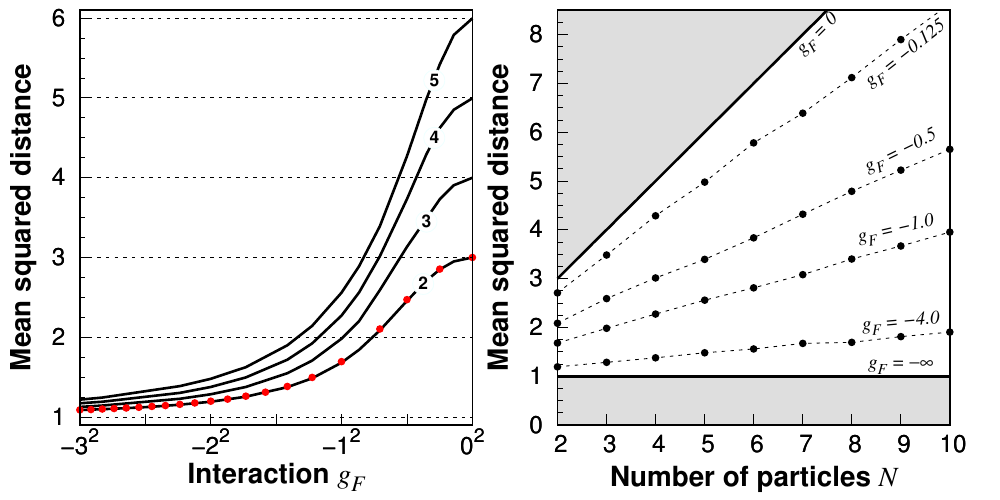}
\caption{The mean squared distance $\sigma^2$ between two fermions defined by Eq. (\ref{MSD}) as a function of interactions $g_F$ for different number of particles {\bf (left panel)} and as a function of the number of particles $N$ for some exemplary interactions {\bf (right panel)}. For comparison, the exact results obtained for $N=2$ case are marked by red dots. In the two limiting cases $g_F=0$ and $g_F\rightarrow -\infty$, the mean squared distance can be determined analytically and it is equal to $N+1$ and $1$, respectively. Note that for better visibility, on the left panel we use a nonlinear scale for interactions. \label{Fig6}}
\end{figure}
The dependence of the two-particle correlations on interactions and the number of particles can be also visualized when the mean squared distance between two fermions is considered. It is defined directly as
\begin{equation} \label{MSD}
\sigma^2 = \int\!\mathrm{d}x_1\mathrm{d}x_2\, (x_1-x_2)^2 n^{(2)}(x_1,x_2).
\end{equation} 
The quantity reflects the spreading of the two-particle density profile in the position domain and its value is known for any number of particles $N$ in two extreme limits. It is equal to $N+1$ for the non-interacting system ($g_F=0$) while for infinite attractions ($g_F\rightarrow -\infty$) it is independent on $N$ and equal exactly to $1$. For intermediate interactions, the mean squared distance $\sigma^2$ can be quite easily determined by the variational method proposed. In Fig.~\ref{Fig6} we display the results obtained for a different number of particles up to $N=10$ and a whole range of attractions. It is clear, that for all number of particles considered, the squared distance $\sigma^2$ rapidly decreases with attractions and it quickly achieves its asymptotic value $1$. Even for a quite large number of particles, $N=10$, the squared distance $\sigma^2$ is less than $2$ if interactions are not weaker than $g_F=-4.0$. This observation supports our previous single-particle conclusions that different properties of the system with infinite attractions are revealed already for intermediate interactions. 

\begin{figure}
\centering
\includegraphics[width=0.6\linewidth]{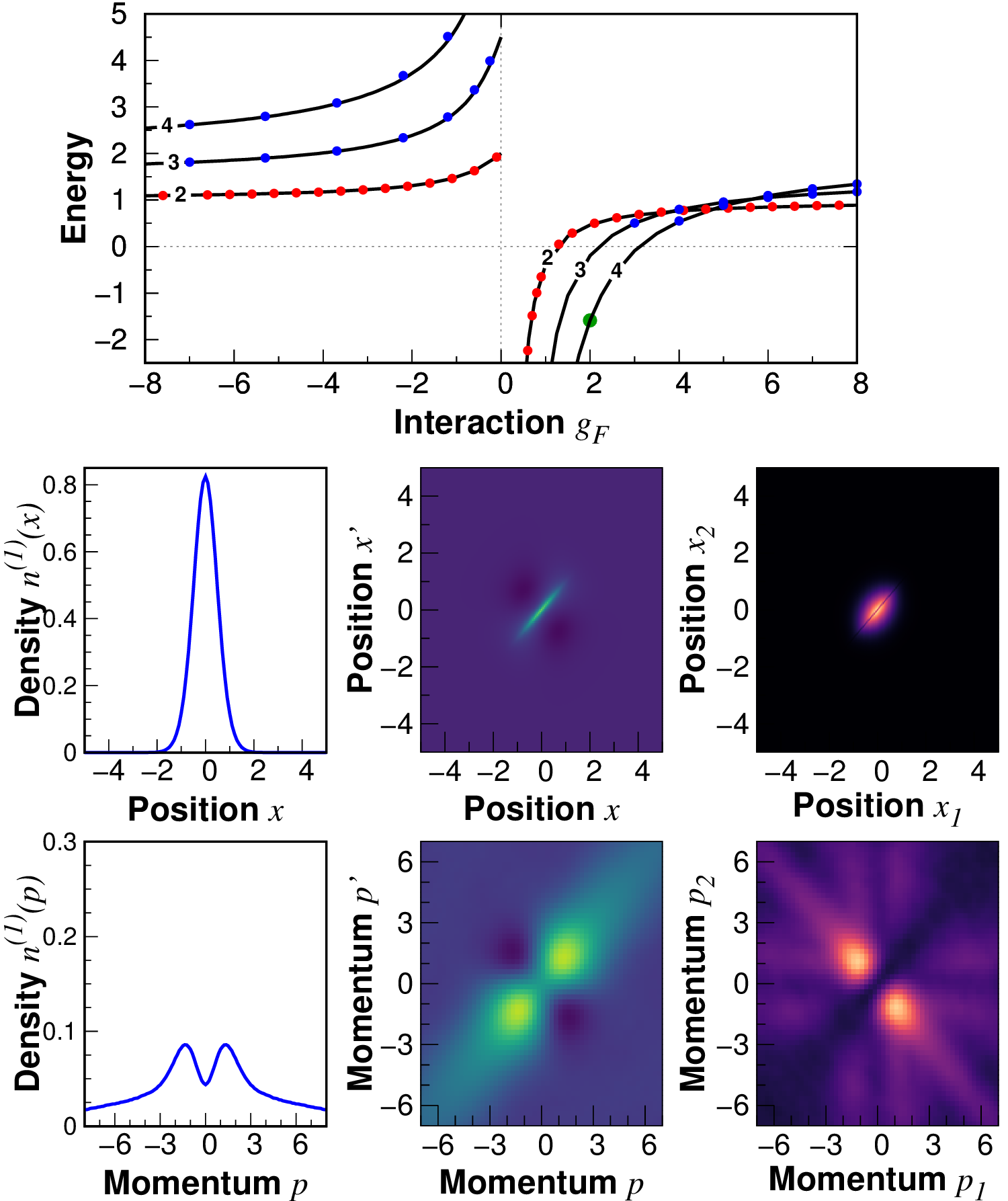}
\caption{{\bf (top panel)} Ground-state energy of $N=2,\ldots,4$ $p$-wave fermions as a function of interactions $g_F$ in the vicinity of the non-interacting system ($g_F=0$). Note clear discontinuity and divergence of the energy when the system is tuned to the non-interacting limit from the repulsive branch. For comparison, with red dots, we mark the exact ground-state energies for $N=2$ calculated analytically as presented in \cite{2006SunPRA}. Blue dots denote corresponding ground-state energies obtained numerically via exact diagonalization of mapped attractively interacting $s$-wave bosonic system of $N=3$ and $N=4$ particles. The green dot denotes the case of $N=4$ fermions and $g_F=2$ which is on the border of the exact diagonalization method achievability but it is well-captured by the variational ansatz. {\bf (bottom panel)} Single- and two-particle properties of the system with $N=4$ fermions in position (upper row) and momentum (bottom row) domain as predicted by the variational ansatz for $g_F=2$ (the case denoted by the green dot in the top panel). Consecutive columns present respectively: single-particle density profiles ($n^{(1)}(x)$ and $\tilde{n}^{(1)}(p)$), single-particle density matrices ($\rho^{(1)}(x,x^\prime)$ and $\tilde\rho^{(1)}(p,p^\prime)$), and two-particle density profiles ($n^{(2)}(x_{1},x_{2})$ and $\tilde{n}^{(2)}(p_1,p_2)$). \label{Fig7}}
\end{figure}

\section{Repulsive $p$-wave interactions}
Up to now in our work we focused mostly on the case of attractively interacting $p$-wave fermions. However, the variational ansatz proposed can be used also to predict different properties of the repulsively interacting system. At this point one should remember however that the $p$-wave interacting system on the repulsive branch is not well-defined in the limit of vanishing interactions. It is clearly visible when the ground-state energy of the system is plotted (top panel in Fig.~\ref{Fig7}). In this limit, the energy is divergent and drops to minus infinity for any number of particles. This quite counterintuitive property of the system is in full agreement with the mapping argument to the bosonic system mentioned in Section~\ref{SectionVariational}. Indeed, a weakly repulsive system of $p$-wave fermions corresponds to the strongly attractive $s$-wave bosons for which energy obviously drops to minus infinity along with increasing attractions. This observation simply means that any properties of a weakly repulsive system cannot be determined with any perturbative analysis starting from the non-interacting system. In this regime, the mentioned mapping from the corresponding bosonic system also does not provide any solution since determining the ground-state wave function for a strongly attractive bosonic system is in practice not possible. The variational approach proposed does not have these limitations and appropriately captures the ground-state properties in a whole range of interactions. As an example, in Fig.~\ref{Fig7}, we display the ground-state energy for $N=2,\ldots,4$ as predicted by the ansatz (top panel). For clarity and credibility, in the case of $N=2$, we compare predicted energies with analytically exact ones (red dots). Additionally, for $N=3$ and $N=4$ we also show ground-state energies obtained via mapping from exactly diagonalized weakly attractive bosons (blue dots) resulting for quite strong $p$-wave repulsions. Note almost perfect compatibility of these results with predictions of the variational ansatz. In the bottom panel we show different single- and two-particle properties of the system of $N=4$ fermions and interaction strength $g_F=2$ which is beyond accuracy of any reasonable numerical approach.  
 
\section{Conclusions}
We have shown that the many-body ground-state of interacting $p$-wave fermions confined in a one-dimensional harmonic trap can be well approximated by a simple variational wave function of the Jastrow type. However, in contrast to the original idea of Jastrow, instead of utilizing a known analytical solution of the corresponding two-body problem, we propose (similarly as we did previously for $s$-wave interactions in \cite{2018KoscikEPL}) to use a much simpler correlated pair function which takes inter-particle interactions precisely into account but significantly simplifies numerical calculations. In this way, we were able to determine different single- and two-particle properties (in the position as well as in the momentum domain) of the system containing up to $10$ particles in a whole range of attractive interactions. We have also briefly discussed the repulsive branch of interactions showing that the variational ansatz appropriately describes the system in a whole range of interactions. It is particularly important when weak repulsions are considered since then all the methods based on perturbative arguments break down.

It is worth pointing out that the scheme proposed is very flexible and general since the contact condition (\ref{condition}) is included in the trial wave function independently on the external confinement. Therefore, by simple modifications of the analytical part in (\ref{jastrow}) one can repeat the scheme for other one-dimensional traps. Moreover, whenever the accuracy of the approximation is insufficient, the same modification can be exploited to propose an another, more adequate family of variational trial functions without modifying the pair-correlation part. Finally, since a whole strategy relies on an appropriate inclusion of the contact condition, a very similar scheme can be used for zero-range forces other than $p$-wave.

\section{Acknowledgments}
This work was supported by the (Polish) National Science Center Grant No. 2016/22/E/ST2/00555.

\appendix

\section{Results for $N=3$ and $N=5$}
\begin{figure}[!]
\centering
\includegraphics[width=0.7\linewidth]{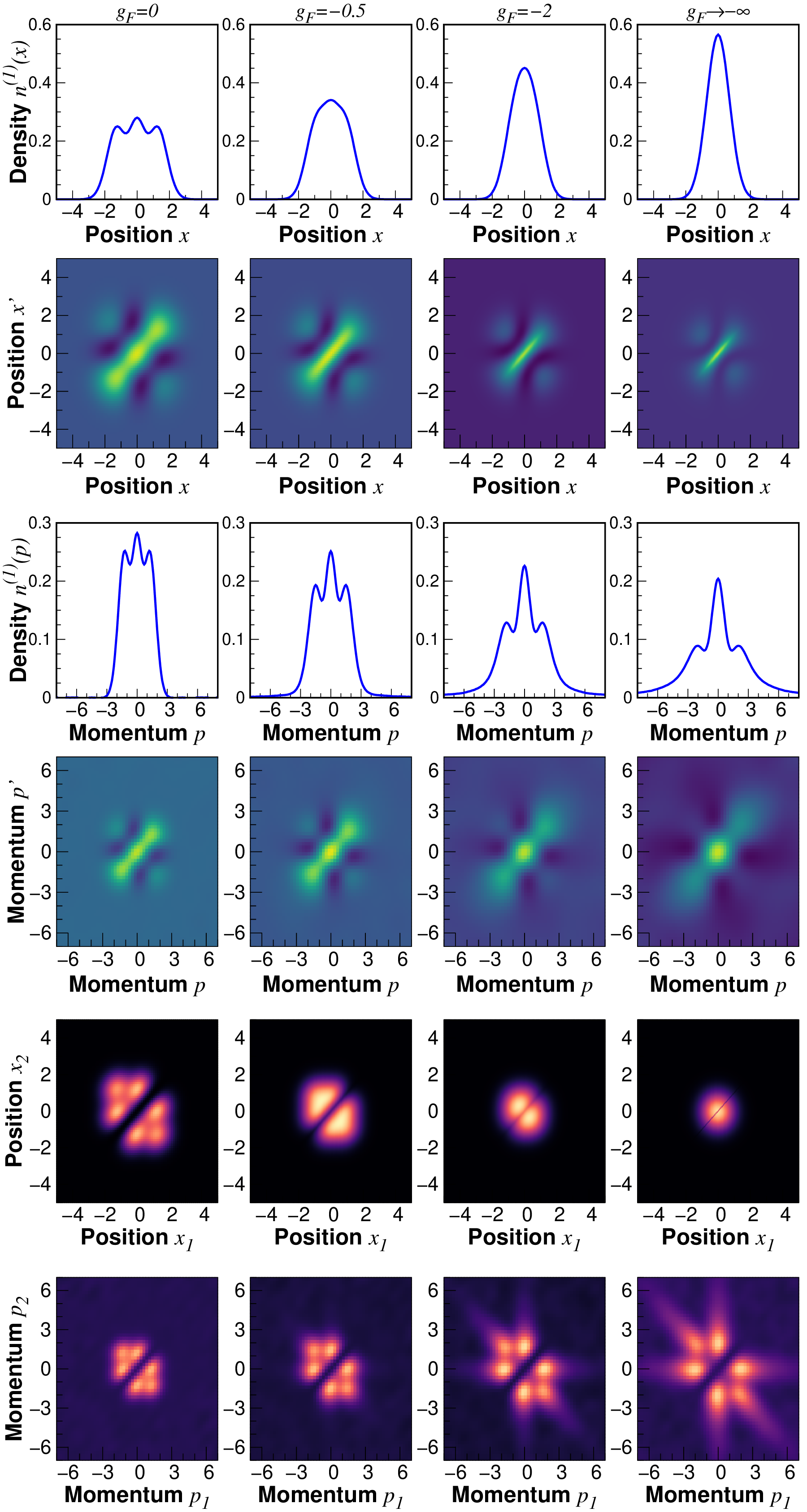}
\caption{Single-particle {\bf (top four rows)} and two-particle {\bf (two bottom rows)} properties of the system of $N=3$ fermions and different interactions strengths. Figure should be compared with the corresponding Fig.~\ref{Fig2} and Fig.~\ref{Fig5} presented for $N=4$ in the main text. \label{FigA1} }
\end{figure}

\begin{figure}[!]
\centering
\includegraphics[width=0.7\linewidth]{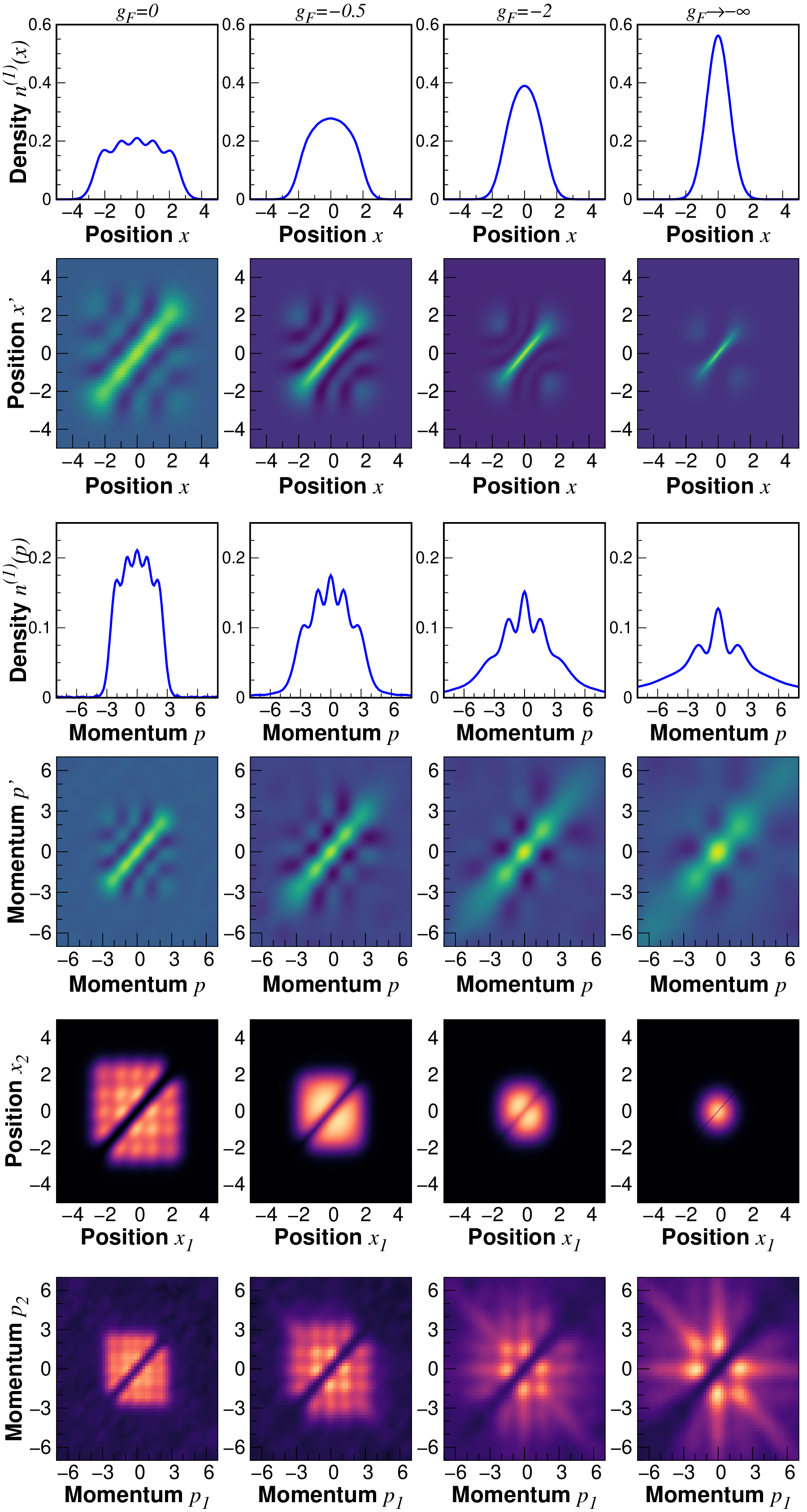}
\caption{Single-particle {\bf (top four rows)} and two-particle {\bf (two bottom rows)} properties of the system of $N=5$ fermions and different interactions strengths. Figure should be compared with the corresponding Fig.~\ref{Fig2} and Fig.~\ref{Fig5} presented for $N=4$ in the main text. \label{FigA2} }
\end{figure}

\begin{figure}
\centering
\includegraphics[width=0.7\linewidth]{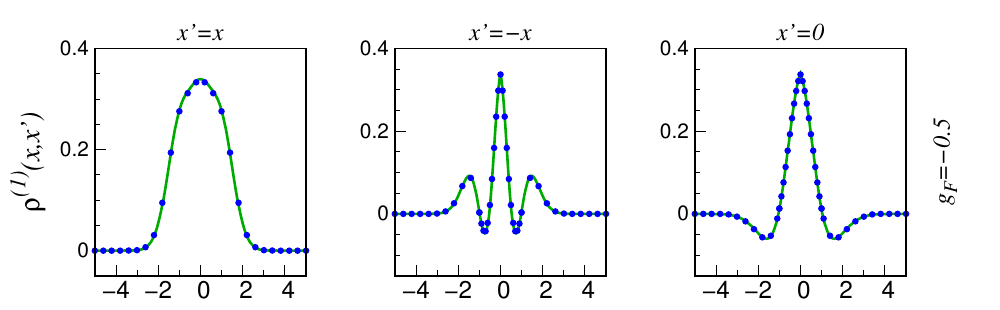}
\includegraphics[width=0.7\linewidth]{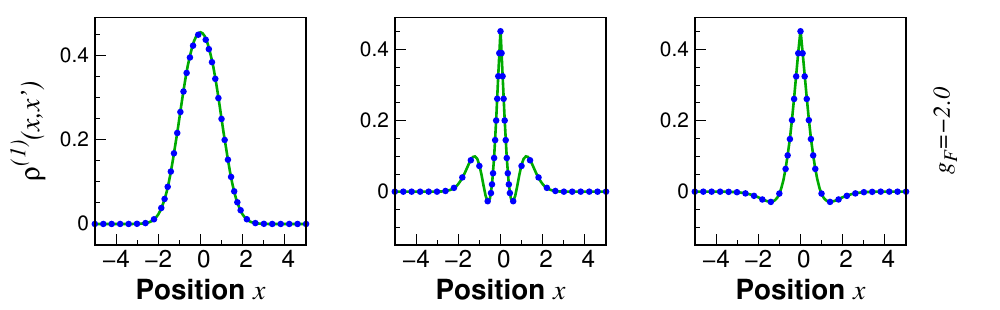}
\caption{Comparison of the single-particle reduced density matrices $\rho^{(1)}(x,x')$ obtained for $N=3$ fermions from the variational ansatz approach (solid green lines) and obtained numerically via exact diagonalization (blue dots). Successive columns correspond to cuts of the density matrix along three different directions: diagonal ($x'=x$), off-diagonal ($x'=-x$), and horizontal ($x'=0$). Top and bottom  row correspond to two different interaction strengths $g_F=-0.5$ and $g_F=-2.0$, respectively. Figure should be compared with the corresponding Fig.~\ref{Fig3} presented for $N=4$ in the main text.\label{FigA3}}
\end{figure}

For completeness of the discussion in this Appendix we present results obtained for different numbers of particles. In Fig.~\ref{FigA1} and Fig.~\ref{FigA2} we display the same quantities as shown in Fig.~\ref{Fig2} and Fig.~\ref{Fig5} but for $N=3$ and $N=5$ particles, respectively. In Fig.~\ref{FigA3}, correspondingly to Fig.~\ref{Fig3} in the main text, we present comparison of predictions served by the variational ansatz and the exact diagonalization for the reduced single-particle density matrix for $N=3$ fermions.

\newpage
\section*{References}
\bibliographystyle{iopart-num}
\bibliography{biblio}

\end{document}